\documentclass{aa}

\usepackage{graphicx}

\input{epsf.sty}

\begin{document}

   \thesaurus{06     
              (02.01.2;  
               04.19.1;  
               08.14.2;  
               08.15.1;  
               08.09.2 HS\,1449+6415)} 
   \title{A Newly Discovered SU\,UMa-Type Dwarf Nova, HS\,1449+6415}

   \subtitle{}

   \author{D. Nogami\inst{1}\thanks{Visiting Astronomer, German-Spanish
Astronomical Center, Calar Alto, operated by the Max-Planck-Institut
f\"{u}r Astronomie, Heidelberg, jointly with the Spanich National
Commission for Astronomy.}, D. Engels\inst{2}, B.T. G\"{a}nsicke\inst{1},
E.P. Pavlenko\inst{3, 4}, R. Nov\'{a}k\inst{5} and K. Reinsch\inst{1}}

   \offprints{D. Nogami (daisaku@uni-sw.gwdg.de)}

   \institute{Universt\"{a}ts-Sternwarte, Geismarlandstr. 11, D-37083
                 G\"{o}ttingen, Germany
                 \and
              Hamburger Sternwarte, Universt\"{a}t Hamburg,
                 Gojenbergsweg 112, D-21029 Hamburg, Germany
                 \and
              Crimean Astrophysical Observatory, p/o Nauchny,
                 98409 Crimea, Ukraine
                 \and
              Isaac Newton Institute Chile, Crimean Branch
                 \and
              Nicholas Copernicus Observatory, Krav\'{\i} hora 2,
                 Brno 616 00, Czech Republic
             }
   \date{Received; accepted}

\authorrunning{Nogami et al.}
\titlerunning{HS 1449+6415}

   \maketitle

   \begin{abstract}

We report time-resolved photometric observations of the Hamburg
Quasar Survey-selected dwarf nova HS\,1449+6415 (= RX\,J1450.5+6403)
during the interval from JD 2451672 to 2451724.  During a long
outburst with a duration of $\sim$12 days, we detected superhumps
with a period of 0$\fd$0601(5), revealing that this star is an
SU\,UMa star.  A short outburst which lasted 3 days was found about
40 days after the onset of the superoutburst.  From quiescence
observations, we determine an orbital period of 0$\fd$05898(2).
The small outburst amplitude of $\sim$3.5 mag, the short orbital
period, and the normal outburst cycle length suggest that
HS\,1449+6415 is an intermediate object between ER\,UMa stars
with high mass-transfer rates and WZ\,Sge stars with very low
mass-transfer rates.

   \keywords{Accretion disks -- Surveys -- Cataclysmic variables
                -- Stars: oscillation --
                Stars: individual: HS\,1449+6415}
\end{abstract}

%

\section{Introduction}

Dwarf novae are a sub-class of the cataclysmic variable stars (CVs),
which show quasi-periodic outbursts (for a review, see Warner
\cite{w95}).  SU\,UMa-type dwarf novae show two types of outbursts,
namely normal outbursts, and so-called superoutbursts.  Normal
outbursts (typically 2--4 mag) last 3--5 days, and superoutbursts
(0.5--1.0 mag brighter than normal outbursts) last about 3--4 times
longer than normal outbursts.  Between two successive superoutbursts,
3--10 normal outbursts occur.  This behavior is now well explained
in the thermal-tidal disk instability model (e.g. Osaki \cite{o96}):
a normal outburst is caused by a thermal limit-cycle based on the
strong dependence of the hydrogen opacity on the ionization state.
in the accretion disk.  When the disk has enough mass to extend
itself radially to the critical tidal radius at the maximum of a
normal outburst, efficient removal of angular momentum due to tidal
stress induced by the secondary maintains the disk longer in the
hot state (superoutburst).  The tidal effects force the accretion
disk into a slowly precessing eccentric form.  The beat phenomenon
of this disk precession and the orbital motion yields photometric
variations with a period slightly longer than the orbital period
$P_{\rm orb}$ (e.g. Whitehurst \cite{w88}, \cite{w94}).  Such
oscillations called superhumps are observed only during
superoutbursts.

Some SU\,UMa-type systems with short orbital periods ($\sim$ 80--90
min) show various patterns of the outburst and the superhump
evolution which are not easily explained in the current disk
instability scheme, such as stars of the WZ\,Sge-subclass (see e.g.
Nogami \cite{n98}) and the ER\,UMa-subclass (see e.g. Kato et al.
\cite{k98}).  Although several models have been proposed to explain
their outburst behaviors, especially that of WZ\,Sge stars, a
common understanding has not yet been reached.  This is partly
because there are few observations of dwarf novae just before a
superoutburst, owing to the difficulty of predicting a superoutburst.

In this Paper, we report photometric observations of HS\,1449+6415
(= RX\,J1450.5+6403), which has been identified as a CV candidate by
Bade et al. (\cite{b98}) based on its emission-line spectrum in the
Hamburg Quasar Survey (HQS) and on the X-ray emission in the ROSAT
All Sky Survey.  The CV nature of HS\,1449+6415 has been onfirmed by
spectroscopic observations (Jiang et al. \cite{j00}).  The HQS has
lead to discoveries of some interesting dwarf novae, for instance,
the eclipsing stars EX Dra (= HS\,1804+6753, Fiedler et al.
\cite{f97}, and references therein) and HS\,0907+1902 (G\"{a}nsicke
et al. \cite{g00}).  HS\,1449+6415 is also identified with
USNO--A2.0 1500--05846206: r\_mag = 16.3, b\_mag = 15.8; $\alpha$ =
14$^{\rm h}$ 50$^{\rm m}$ 38$\fs$3, $\delta$ = +64$^{\circ}$ 03$'$
29$''$ (J2000.0). Skiff (\cite{s00}) pointed out another
identification as FBS\,1449+642, a blue stellar object found in the
First Byurakan Spectral Sky Survey (Abrahamian \& Mickaelian
\cite{a94}).

\section{Observations and Results}

\begin{table*}
   \caption[]{Observation log.}
      \begin{tabular}{c c c c c c c}
            \hline
           Coverage   & Int. time & Number of  & Filter & Average, & State & Observatory$^1$ \\
       HJD$-$2450000  & (s)     & observations & & RMS mag \\
       \hline
       1672.394 -- 1672.437   &  60   &  46  & $V$ & 17.3($\pm$ 0.2) & quiescence & CA \\
       1675.437 -- 1675.668   &  60   & 145  & $V$ & 17.5($\pm$ 0.2) & quiescence & CA \\
       1676.604 -- 1676.661   &  60   &  19  & $V$ & 17.6($\pm$ 0.3)& quiescence & CA \\
       1680.344 -- 1680.603   &  50   & 407  & $R_{\rm c}$ & 14.2($\pm$ 0.1) & superoutburst & NC \\
       1687.320 -- 1687.466   & 120   &  87  & $R_{\rm c}$ & 15.1($\pm$ 0.1) & superoutburst & CR \\
       1691.351 -- 1691.460   & 250   &  33  & $R_{\rm c}$ & 16.6($\pm$ 0.1) & decline & CR \\
       1716.366 -- 1716.429   & 180   &  26  & $R_{\rm c}$ & 17.3($\pm$ 0.2) & quiescence & CR  \\
       1719.364 -- 1719.467   & 180   &  45  & no filter & 17.4$^2$($\pm$ 0.2) & quiescence & CR \\
       1720.308 -- 1720.453   & 180   &  58  & no filter & 17.5$^2$($\pm$ 0.2) & quiescence & CR \\
       1721.295 -- 1721.457   &  90   & 112  & $R_{\rm c}$ & 15.2($\pm$ 0.1) & normal outburst & CR \\
       1722.299 -- 1722.433   & 120   &  31  & $R_{\rm c}$ & 15.7($\pm$ 0.1) & normal outburst & CR \\
       1724.298 -- 1724.355   & 120   &  28  & $R_{\rm c}$ & 16.8($\pm$ 0.2) & normal outburst & CR \\
            \hline
      \end{tabular}

$^1$CA = Calar Alto 1.23m, NC = Nicholas Copernicus Observatory 40cm,
CR = Crimean Astrophysical Observatory 38cm

$^2$These magnitudes are deduced using the $R_{\rm c}$ magnitude of
the comparison star.
\end{table*}

\begin{figure}[t] 
\includegraphics[width=8.8cm]{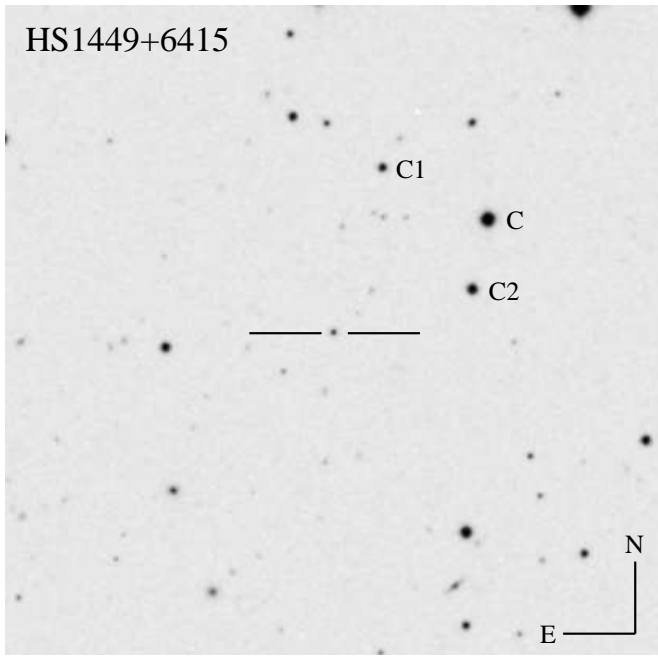}
\caption[]{Finding chart of HS\,1449+6415 obtained from the Digitized
Sky Survey.  The field of view is $7' \times 7'$. The star `C' is the
comparison star, and the stars `C1' and `C2' are the check stars.}
\end{figure}

In order to measure the orbital period, we carried out time-resolved
photometry of HS\,1449+6415 using a standard Johnson $V$ filter and
a TEK CCD attached at the Cassegrain focus of the 1.23-m telescope
at the Calar Alto Observatory (CA) on 2000 May 7, 10, and 11, when
the star was in quiescence.  The exposure time was 60 sec, and the
dead time between exposures was 12 sec, adopting the fast read-out
mode and limiting the read-out region on the CCD.  The journal of
our observations is shown in Table 1.

The first outburst of this star was reported to the VSNET by
Vanmunster (\cite{v00a}), three days after our CA observations.  We
immediately started a photometric run at the Nicholas Copernicus
Observatory (NC) on 2000 May 15.  The NC data were obtained through
a Kron-Cousins $R_{\rm c}$-band filter with an SBIG ST-7 camera and
a 40cm Newtonian telescope.  The exposure time and the dead time
were 50 s and typically 5 sec, respectively.

At the Crimean Astrophysical Observatory (CR), we then monitored
the behavior of HS\,1449+6415 during the late stages of the long
outburst.  The CR observations were carried out using a 38-cm
Cassegrain telescope with an SBIG ST-7 camera and with a
Kron-Cousins $R_{\rm c}$ or no filter, depending on the atmospheric
condition and the brightness of the target.  The exposure time was
also varied between 90 s and 250 s.  This outburst proved to be in
the final decline during the second run at CR, implying that the
outburst lasted $\sim$12 days.  

After the standard reduction of the frames,  we measured the
magnitudes of the variable using either an aperture photometry
routine in MIDAS (CA data), the CCD photometry package Munidos
(http://ian.cz/munipack), which is based on DAOPHOT (Stetson
\cite{s87}) (NC data), or an aperture photometry package developed
by V. P. Goransky (CR data).

To calibrate the magnitudes, we used the local comparison star `C'
(= USNO--A2.0 1500--05845729; Fig. 1), whose photometric properties
were given by Henden (\cite{h00})\footnote{His original BVRI
photometry file for this star is available at
ftp://ftp.nofs.navy.mil/pub/outgoing/aah/sequence/j1450.dat.} as
$V$ = 13.151(17), $B-V$ = 0.701(5), and $V-R_{\rm c}$ = 0.384(9).
Constancy of the comparison star within 0.02 mag during our run was
checked using stars `C1' and `C2'.

\begin{figure}[t] 
\includegraphics[width=8.8cm]{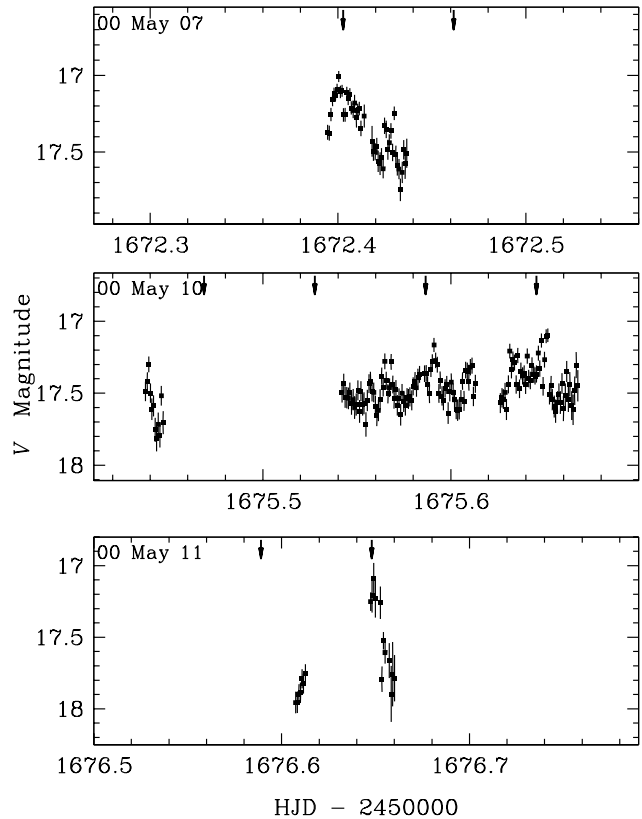}
\caption[]{CA Light curves of HS\,1449+6415 in quiescence.  The
arrows indicate times at $\phi_{\rm phot}$ = 0.0, which corresponds
to the phase in Fig. 3.}
\end{figure}

\begin{figure}[t] 
\includegraphics[width=8.8cm]{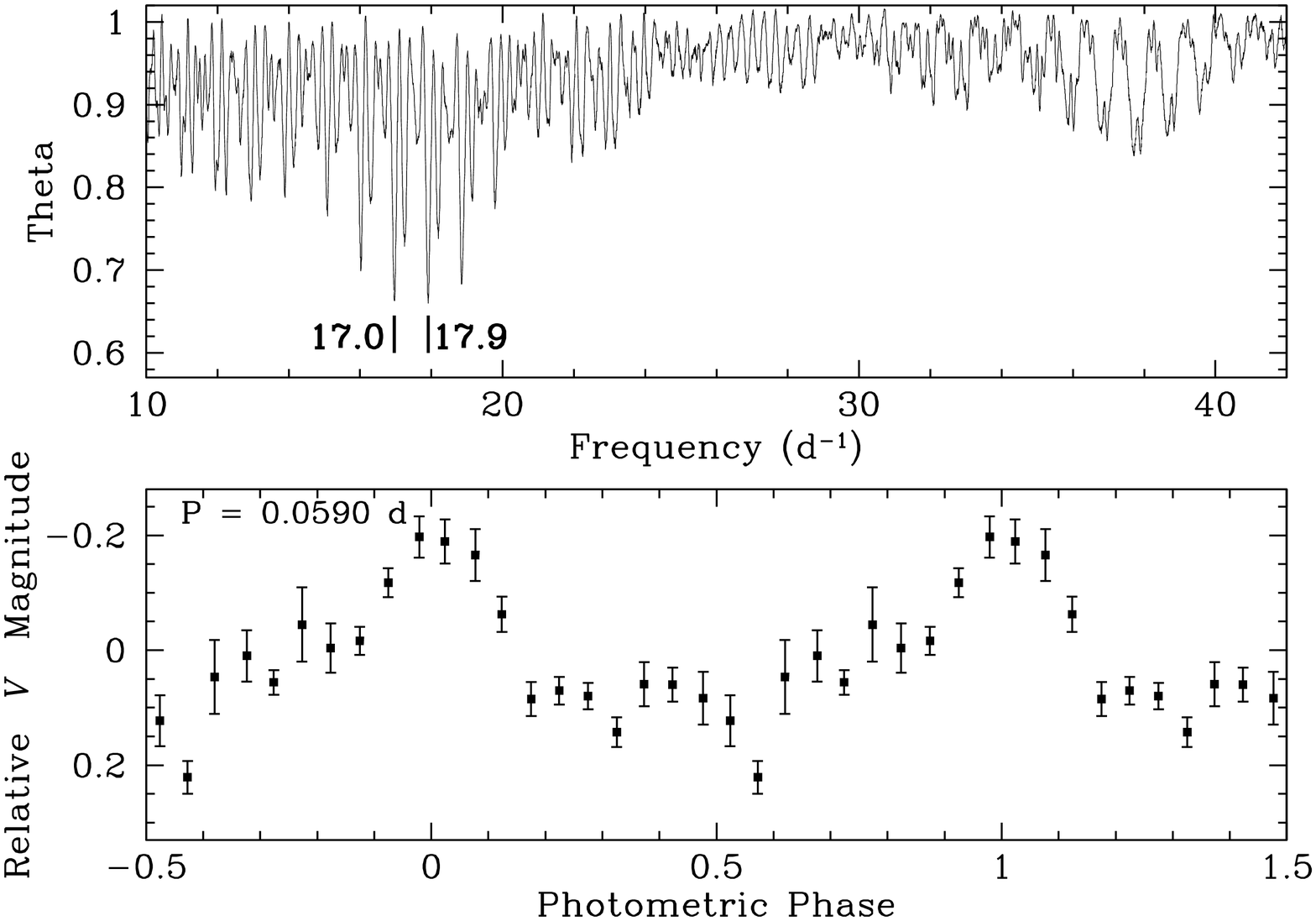}
\caption[]{Upper panel: Theta diagram of the period analysis for the
CR data in Fig. 2 using the PDM method.  There are two candidates for
the orbital period: 0$\fd$0558(2) (17.9 d$^{-1}$) and 0$\fd$0590(2)
(17.0 d$^{-1}$).  Lower panel: Light curve folded by the period of
0$\fd$0590.}
\end{figure}

HS\,1449+6415 was in quiescence during the CA observations (Table 1).
The light curves in quiescence are shown in Fig. 2; large amplitude
flickering seen each night with no hint of an eclipse.  After
subtraction of nightly averages from the quiescent CA data, a period
analysis using the Phase Dispersion Minimization (PDM) method
(Stellingwerf \cite{s78}) yields two candidate periods, 0$\fd$0558(2)
(= 17.9 d$^{-1}$, 80.4 min) and 0$\fd$0590(2) (= 17.0 d$^{-1}$, 85.0
min) as shown in Fig. 3, which are one-day aliases of each other.
We can not distinguish between these candidates from our CA data
alone because of the short coverage.

Time resolved photometry obtained at NC on May 15 (Fig. 4) and 22
during the long outburst clearly revealed the existence of
superhumps, proving the SU\,UMa nature of HS\,1449+6415.  These
superhump data help to determine the true orbital period.  After
subtraction of nightly averages from the data, we performed a PDM
period analysis.  The two best candidates for the superhump period
$P_{\rm sh}$ are 0$\fd$06019(2) (= 16.614(5) d$^{-1}$, 86.67 min)
and its 7-day alias, 0$\fd$05965(2) (= 16.764(5) d$^{-1}$, 85.90
min).  Nevertheless, the latter possibility can be rejected from
Vanmunster et al. (\cite{vsf00}), who have much denser coverage of
this superoutburst than we have. The superhump period is, therefore,
determined to be 0$\fd$06019(2), which exactly accords with the
value in Vanmunster et al. (\cite{vsf00}).  The superhump excess
$\epsilon$ (= $(P_{\rm sh} - P_{\rm orb})/P_{\rm orb}$) is
7.9\, \% and 2.0\,\% for the two $P_{\rm orb}$ candidates of
0$\fd$0558 and  0$\fd$0590, respectively.  Since SU\,UMa stars with
short orbital periods have $\epsilon$ $\simeq$ 1--3\,\% (see e.g.
Nogami et al. \cite{nmk97}), this well-determined superhump period
strongly supports the latter $P_{\rm orb}$ candidate.  Moreover,
Thorstensen (\cite{t00}) reported two orbital period candidates of
0$\fd$0588 and 0$\fd$0599 deduced from radial velocity measurements.
The former is in good agreement with our candidate period of
0$\fd$0590(2).  Hence, we can safely regard this period as the
orbital period of HS\,1449+6415.  The lower panel of Fig. 3 shows
the light curve obtained by folding all the CA data with the orbital
period.  The phase was determined to place the large peak at $\phi$
= 0.0.

\begin{figure}[t] 
\includegraphics[width=8.8cm]{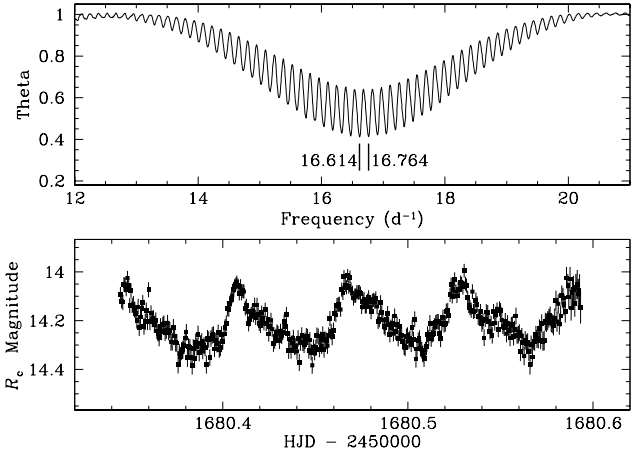}
\caption[]{Fully grown superhump seen on 2000 May 15 (NC data,
upper panel).  A PDM period analysis yields two candidates of the
superhump period of 0$\fd$06019(2) and 0$\fd$05965(2) (lower panel),
but the latter is rejected from Vanmunster et al. (\cite{vsf00}).}
\end{figure}

To check whether our periodogram may be biased by night-to-night
variability, we made another period analysis using only the
longest continuous observing sequence obtained on May 10.  The
resultant theta diagram yields two broad peaks around frequencies
of $\sim$18.0 d$^{\rm -1}$ and $\sim$38 d$^{-1}$  with almost the
same significance.  This may indicate the underlying double-peaked
form of the orbital light curve found by Kato et al. (\cite{k00})
in observations obtained during quiescence on 2000 February 24 at
the Nyrola Observatory.  Their period analysis has yielded two
strong peaks at 0$\fd$0289 and its double.  Double-peaked orbital
modulations in quiescence are found in WZ\,Sge (e.g. Warner \&
Nather \cite{wn72}) and AL\,Com (see e.g. Szkody et al. \cite{s89}).

\begin{figure}[t] 
\includegraphics[width=8.8cm]{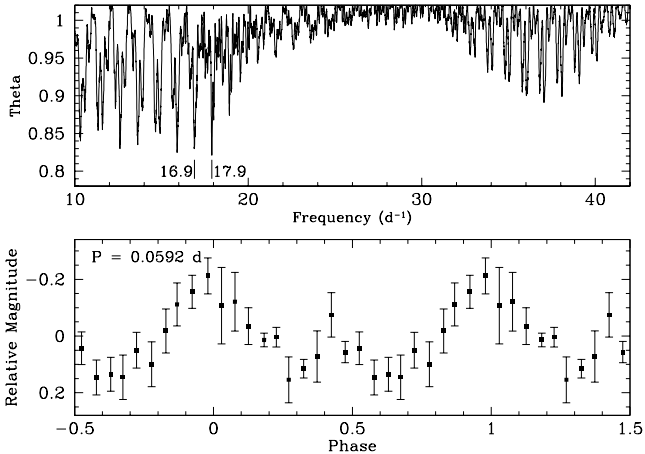}
\caption[]{Upper panel: Theta diagram for the CR data obtained
between JD 2451716 and JD 245120 in quiescence.  There are also
peaks at 0$\fd$0559(4) (17.9 d$^{-1}$) and 0$\fd$0592(4)
(16.9 d$^{-1}$), in good agreement with the result of Fig. 3.
Lower panel: Light curve folded by the period of 0$\fd$0592 and 
howing a double-peaked shape.}
\end{figure}

Quiescent data was also obtained at CR around JD 2451716.  Fig. 5
exhibits the result of a PDM period analysis after subtraction of
nightly averages and the folded light curve of the data from JD
2451716 to 2451720.  The shape of the orbital modulation is almost
the same as in Fig. 3 (for instance, the full amplitude of $\sim$0.4
mag) but the secondary peak around $\phi$ = 0.4 is stronger.

A PDM period analysis using all our quiescence data obtained at
CA and CR yields the refined photometric ephemeris:
\begin{equation}
HJD_{\rm max} = 2451672.047(2) + 0.05898(2) \times E \,.
\end{equation}
The superhump maxima in Fig. 4 are around $\phi$ = 0.77 in this
ephemeris.

\begin{figure}[t] 
\includegraphics[width=8.8cm]{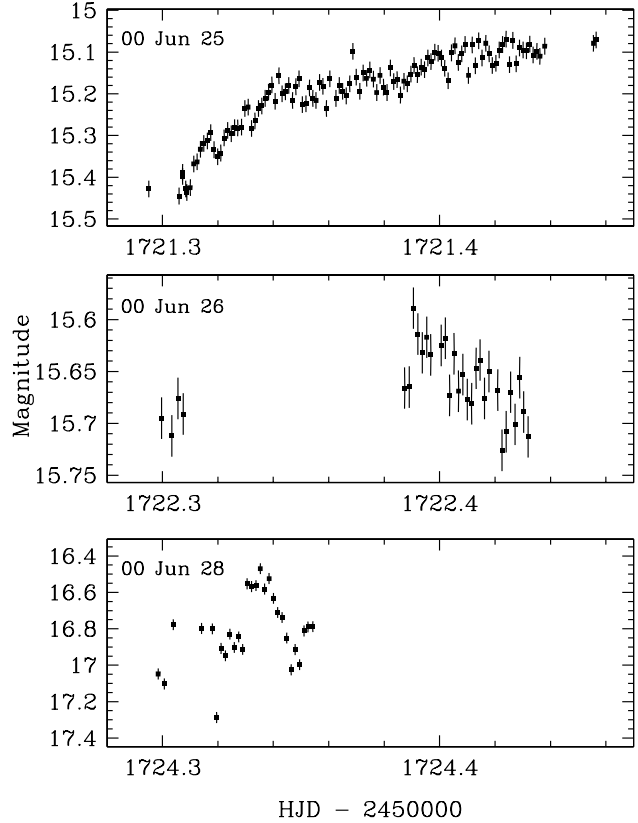}
\caption[]{Light curves during a normal outburst lasting only 3
days.  While no periodic variations and only weak flickering are
seen on June 25 and 26, flickering of $\sim$ 0.5 mag on June 28
when HS\,1449+6415 is about 0.6 mag brighter than in quiescence.}
\end{figure}

A short outburst lasting only 3 days were caught on JD 2451721 in
the course of our monitoring of this star at CR (Fig. 6).  This
duration is typical for a normal outburst of SU\,UMa stars.  There
are no periodic variations but weak flickering in the first and
the second night of this outburst.  Large flickering of $\sim$ 0.5
mag can be seen on HJD 2451724, when this variable star was still
about 0.5 mag brighter than its quiescence level.

\section{Discussion}

The superhump period we derived is 0$\fd$0601(6), 1.9\,\% longer than
$P_{\rm orb}$ of 0$\fd$05898(2).  This orbital period is close to the
minimum of the period distribution of SU\,UMa-type dwarf novae (see
e.g. Nogami et al. \cite{nmk97}).  In general, observations have
shown that the amount of time after the maximum of a superoutburst
needed for SU\,UMa stars to evolve superhumps increases towards
shorter orbital periods (see e.g. Table 1 in Osaki \cite{o96}). This
correlation is also expected on the theoretical grounds: Lubow
(\cite{l91}) showed that the growth rate of the eccentricity is
proportional to $q^2$, where $q$ is the mass ratio of the secondary
mass to the primary mass (= $M_{\rm 2}/M_{\rm 1}$).  Thus, if a
normal SU\,UMa star, superhumps in HS\,1449+6415 are expected to
appear at least 4 days after the supermaximum.  In contrast, fully
grown superhumps with an amplitude of 0.32 mag were detected less
than 2 days after the maximum of the superoutburst detected on 2000
May 14.864 (UT), only 2.7 day after the end of our CA observations
(Vanmunster \cite{v00b}; Vanmunster et al. \cite{vsf00}), taking
into account the rise time to the supermaximum.  A determination of
the binary parameters of HS\,1449+6415 by spectroscopy may thus be
able to test the tidal instability theory.

Kato et al. (\cite{k98}) pointed out that all ER\,UMa stars, i.e.,
the most active SU\,UMa-type dwarf novae, exhibit a peculiar
phenomenon: the amplitude of the superhump is largest around the
superoutburst maximum and decreases as the time proceeds even though
the superhump period remains almost constant (see also Kato et al.
\cite{k96}).  ER\,UMa stars also have very short $P_{\rm orb}$
close to that of HS\,1449+6415.  The early development of the
superhump in the present system may be related to the superhump
evolution in ER\,UMa stars which is still theoretically unexplained.
To trace the variation of the superhump in HS\,1449+6415 throughout
a superoutburst is another future work which should be performed.

WZ\,Sge stars, another extreme subset of SU\,UMa-type dwarf novae
(O'Donoghue et al. \cite{o91}, and references therein) also have
very short orbital periods, similar to that of HS\,1449+6415.
However, WZ\,Sge stars shows only superoutbursts with a very long
recurrence time ($\geq$ 10 yr) and with a very large amplitude
($\Delta V\geq7$), but no normal outbursts.  For this behavior,
Meyer-Hofmeister et al. (\cite{m98}) proposed a model in which the
accretion disk steadily grows during quiescence, and the tidal
instability starts before the superoutburst onset.  If this model
were applicable to HS\,1449+6415, we could observe superhumps also
before the onset of a superoutburst.  However, only modulations
with the orbital period were detected in our data.  The amplitude
variation in our data before the superoutburst may indicate a beat
phenomenon of the orbital hump and the superhump, but a much longer
photometric pre-outburst observation would be needed to confirm
this hypothesis.

Note that the average profile of the orbital modulation is almost
the same between just before the superoutburst and before the normal
outburst.  This implies that the physical status of the accretion
disk and the hot spot, which are thought to emit most of the optical
light in quiescence, is quite similar in both these periods.  Nothing
special seemed to occur before the superoutburst, compared to the
situation before the normal outburst, although more observations of
the orbital modulation with much smaller errors are needed to strictly
probe differencies of the accretion disk in different phases.

The brightest magnitude of HS\,1449+6415 reported to VSNET is
$m_{\rm vis}$ = 14.1 (Kinnunen \cite{ki00}), after calibration
using the magnitude of the comparison star (Kinnunen, private
communication) given by Henden (\cite{h00}).  The amplitude of the
superoutburst, thus, is $\sim$3.5 mag, which is the smallest among
SU\,UMa stars with orbital periods shorter than 90 min, other than
ER\,UMa stars (see e.g. Table 1 in Nogami et al. \cite{nmk97}).
As mentioned above, the normal outburst was caught about 40 days
after the onset of the superoutburst.  If this value is considered
as the recurrence cycle of the outbursts, this is typical for an
SU\,UMa star (see e.g. Table 1 in Nogami et al. \cite{nmk97}).
The VSNET
data\footnote{http://www.kusastro.kyoto-u.ac.jp/vsnet/etc/searchobs.html}
can exclude the possibility of an outburst-recurrence cycle shorter
than 14 days.  The combination of the outburst amplitude of 3.5 mag
and the recurrence cycle of 40 days is almost the same as that in
SX\,LMi, which is regarded as the dwarf nova bridging the gap
between ER\,UMa stars and SU\,UMa stars (Nogami et al. \cite{nmk97}).
The orbital period of 0$\fd$06717(11) (= 97 min) of SX\,LMi (Wagner
et al. \cite{w98}) is, however, a little out of range of the orbital
period distribution of ER\,UMa stars and WZ\,Sge stars.  Thus,
HS\,1449+6415 is an intermediate object between ER\,UMa stars and
WZ\,Sge stars, and might be in the evolutionary transition state
between ER\,UMa and WZ\,Sge phases.  HS\,1449+6415 deserves to be
monitored over long timescales to measure the accurate recurrence
cycles of the normal outburst and the superoutburst and to reveal
variations of these cycles and the outburst amplitude.

\begin{acknowledgements}

The authors are really thankful to VSNET contributors for their
observation reports and discussions on VSNET.  Thanks also to
T. Kato and F. V. Hessman for their useful comments.  DN and BTG
were supported by DLR/BMBF grant 50\,OR\,9903\,6.  The HQS was
supported by the Deutsche Forschungsgemeinschaft through grants
Re 353/11 and Re 353/22.

\end{acknowledgements}


\begin{thebibliography}{}

\bibitem[1994]{a94} Abrahamian H. V., Mickaelian A. M., 1994,
Astrofizika 37, 43 (in Russian, translated to English in
Astrophysics 37, 27)

\bibitem[1998]{b98} Bade N., Engels D., Voges W., et al., 1998,
A\&AS 127, 145

\bibitem[1997]{f97} Fiedler H., Barwig H., Mantel K.H., 1997, A\&A
327, 173

\bibitem[2000]{g00} G\"{a}nsicke B.T., Fried R.E., Beuermann K.,
Engels D., Hagen H.-J., Hessman F.V., Nogami D., Reinsch K., 2000,
A\&A 356, L79

\bibitem[1995]{h95} Hagen H.-J., Groote D., Engels D., Reimers D.,
1995, A\&AS 111, 195

\bibitem[2000]{h00} Henden A.H., 2000, vsnet-chat 3339\footnote{The
VSNET publications can be retrieved at
http://www.kusastro.kyoto-u.ac.jp/vsnet/Mail}

\bibitem[2000]{j00} Jiang X.J., Engels D., Wei J.Y., Tesch F.,
Hu J.Y., 2000, A\&A, submitted

\bibitem[1998]{k98} Kato T., Nogami D., Baba H., Masuda S., Matsumoto
K., Kunjaya C., 1998, in {\it Disk Instabilities in Close Binary
Systems}, Mineshige S., Wheeler J.C. (eds.), pp. 45--52 (Tokyo:
Universal Academy Press)

\bibitem[1996]{k96} Kato T., Nogami D., Masuda S., 1996, PASJ 48, L5

\bibitem[2000]{k00} Kato T., Oksanen A., Moilanen M., Kinnunen T.,
2000, vsnet-alert 4858

\bibitem[2000]{ki00} Kinnunen T., 2000, vsnet-obs 27737

\bibitem[1991]{l91} Lubow S. H., 1991, ApJ 381, 259

\bibitem[1998]{m98} Meyer-Hofmeister E., Meyer F., Liu B. F., 1998,
A\&A 339, 507

\bibitem[1998]{n98} Nogami D., 1998, in {\it Disk Instabilities in
Close Binary Systems}, Mineshige S., Wheeler J.C. (eds.), pp. 13--20
(Tokyo: Universal Academy Press)

\bibitem[1997]{nmk97} Nogami D., Masuda S., Kato T., 1997, PASP 109,
1114

\bibitem[1991]{o91} O'Donoghue D., Chen A., Marang F., Mittaz
J.D.P., Winkler H., 1991, MNRAS 250, 363

\bibitem[1996]{o96} Osaki Y., 1996, PASP 108, 39

\bibitem[2000]{s00} Skiff B., 2000, vsnet-alert 4835 

\bibitem[1978]{s78} Stellingwerf R.F., 1978, ApJ 224, 953

\bibitem[1987]{s87} Stetson P.B., 1987, PASP 99, 191

\bibitem[1988]{s89} Szkody P., Howell S.B., Mateo M., Kreidl T.,
1989, PASP 101, 2379

\bibitem[2000]{t00} Thorstensen J.R., 2000, vsnet-alert 4848 

\bibitem[2000a]{v00a} Vanmunster T., 2000a, vsnet-alert 4834 

\bibitem[2000b]{v00b} Vanmunster T., 2000b, vsnet-alert 4838 

\bibitem[2000]{vsf00} Vanmunster T., Skillman D.R., Fried R.E.,
Kemp J., Nov\'{a}k R., 2000, IBVS submitted

\bibitem[1998]{w98} Wagner R.M., Thorstensen J.R., Honeycutt R.K.,
Howell S.B., Kaitchuck R.H., Kreidl T.J., Robertson J.W., Sion E.M.,
Starrfield S.G., 1998, AJ 115, 787

\bibitem[1995]{w95} Warner B., 1995, Cataclysmic Variable Stars
(Cambridge: Cambridge University Press)

\bibitem[1972]{wn72} Warner B., Nather R. E., 1972, MNRAS 156,
297

\bibitem[1988]{w88} Whitehurst R., 1988, MNRAS 232, 35

\bibitem[1994]{w94} Whitehurst R., 1994, MNRAS 266, 35

\end{thebibliography}
\end{document}